\documentclass[aps,twocolumn,showpacs,showkeys,superscriptaddress]{revtex4}
 \usepackage{epsfig,amssymb,bm,graphics,color }
 \usepackage{wrapfig}
\usepackage[caption=false]{subfig}
\usepackage{float}
\def\ba{\begin{eqnarray} }
\def\ea{\end{eqnarray} }
\usepackage{subfig}
\usepackage{epsf}
\usepackage{graphicx}\usepackage{epsfig}
\usepackage{amsmath}
\usepackage{graphics}
 \usepackage{soul,ulem}
 \usepackage{dcolumn}
 \usepackage{marginnote}

\everymath{\displaystyle}
\usepackage{bm}

\usepackage{relsize}

 \def  \Li3 { {\rm {Li}_3}  }
 \def  \li2  { { \rm {Li}_2 } }
\def  \poly4  { { \rm {Li}_4 } }
\def \Imm { \mbox{\rm Im} }
\def \Ree { \mbox{\rm Re} }

\begin{document}
 \title{Some Mathematical Methods Relevant to   Analytical Calculations of  Anomalous Magnetic Moments of Leptons}

\author{ L.P. Kaptari }
\email{kaptari@theor.jinr.ru} \affiliation{Bogoliubov Lab. Theor.
Phys., JINR, Dubna, 141980, Russia}
\author{V.I. Lashkevich}
\email{lashkevich@gstu.gomel.by} \affiliation{Gomel State Technical
University, Gomel, 246746, Belarus}
\author{O.P. Solovtsova}
\email{olsol@theor.jinr.ru}
\affiliation{Bogoliubov Lab. Theor. Phys., JINR, Dubna, 141980,
Russia}
 \affiliation{Gomel State Technical
University, Gomel, 246746, Belarus}

\begin{abstract}

The mathematical methods and approaches used in our analytical
calculations of electromagnetic corrections to the leptonic anomaly
$a_L$, which is caused by a special class of Feynman diagrams with
insertions  of the vacuum polarization operator, are presented and
examined in detail.
The developed methods are essentially  based on the Mellin--Barnes
transform for the propagators of massive particles with  subsequent
application of the PSLQ algorithm to solve the emerging functional
equations. Functional equations are also used in the procedure of
analytic continuation of a particular type of hypergeometric functions
to the entire complex plane of their parameters. It is shown that
these approaches make it possible to obtain explicit expressions for
radiative corrections of rather high orders.
\end{abstract}

\pacs{13.40.Em, 12.20.Ds, 14.60.Ef}

\keywords{lepton anomaly, radiative corrections, high orders,
Mellin--Barnes transform, analytic continuation}

 \maketitle

\section*{Introduction}
Investigation of the nature of the experimentally observed deviation
of the gyromagnetic ratio ($g_L$) of leptons ($L$) from the value
$g_L=2$ predicted by the Dirac equation~\cite{dirac} plays an
important  role in testing  the Standard Model (SM) as a theory of
fundamental interactions, in particular the quantum electro dynamics.
In the literature, it is commonly adopted to   characterize this
deviation by the ratio $a_L\equiv(g_L -2)/2$  which is also known as
the lepton anomaly. Up to now, the measurements of $a_L$ are among the
most precise experiments in particle
physics~\cite{Parker:2018vye,Morel,Fan:2022eto,Fermilab2023}. Accordingly,
these experimental data impose strict  requirements on theoretical
approaches to investigations of the variety of mechanisms contributing
to $a_L$, see e.g., Refs.~\cite{Jegerlehner:2017gek,review-2025}.  The
discussion of the provenance of the $a_L$ anomaly and possible
explanations of the discrepancy between experimental data and
theoretical results obtained with the SM  can be found in
Refs.~\cite{CMD-3:2023alj,CMD-3:2023rfe,Davier-2023-1,Davier-2023-2,Masjuan-2024,Bryzgalov-2024,Volkov-2024}.

Nowadays, theoretical investigations  of the electromagnetic
corrections to $a_L$ originating from Feynman diagrams of high  order
relative to the electric charge $e$
 are performed numerically by special algorithms (see e.g., Refs.~\cite{Laporta:2017okg,elliptic1,Volkov-22,Volkov-2024,Volkov-2023})
allowing for extremely precise numerical calculations.
 It should be noted however,
  that these numerical calculations require quite a lot of computer resources.
  Besides,  a detailed numerical analysis of the contributions of
various mechanisms to $a_L$ is rather hampered. Therefore,  of
particular interest  is the possibility of separating
  a subclass of diagrams that provides explicit  analytical expressions for the anomaly.
Evidently,  in this case a comparison of numerical calculations with
exact analytical results becomes feasible. Consequently, these
comparisons make it possible to  estimate  the reliability  of the
methods and algorithms used in numerical calculations. The so-called
``bubble''\,-like diagrams with insertions of the photon polarization
operator consisting only of closed lepton loops serves as an example
of such class of diagrams. In Ref.~\cite{Aguilar:2008qj},  based on
the Mallin-Barnes (MB) representation for massive propagators, the
contributions of ``bubble''\,-like diagrams were  obtained in
analytical forms and investigated in detail. Within the MB
representation, the corresponding expressions for the anomalous
magnetic moment were
 represented as a series with respect to  the ratio of lepton masses  $r=m_\ell/m_L$, where
 $m_L$ and $m_\ell$ denote the mass of the considered leptons and of leptons in the polarization operator, respectively.
In Ref.~\cite{Solovtsova23}, this approach was applied to derive
analytical expressions for the anomaly $a_L$ for each kind of leptons
$L=e,~\mu$ and  $\tau$  in the whole   interval of the mass ratio
 $0 <\, m_\ell/m_L\, <\, \infty$. It was shown
  that corrections to  $a_L$ from the diagrams
  with insertions of the photon polarization operator can be expressed
  as corrections from diagrams with the exchange of only one, but massive, photon.
 The MB representation for massive propagators enables one to express $a_L$ as
a one-dimensional integral of two Mellin momenta in the complex plane.
It should be noted that the MB transform is a quite general method
widely used in physics of elementary particles, see e.g.,
Ref.~\cite{Mellin-book}. Among the first applications of the method
one can mention Refs.~\cite{Boos,Friot:2005cu} where the method was
used
 to calculate the anomaly of massive diagrams. As for direct applications    of the MB transform in calculations
of  $a_L$, one should note Ref.~\cite{Friot:2005cu}  (see also also
Refs.~\cite{Aguilar:2008qj,Solovtsova23}).

In this article, we  present some mathematical methods and approaches
we
  used in Refs.~\cite{Solovtsova-2019, Solovtsova-Vesti,Solovtsova23,Solovtsova-2024} to
derive   explicit analytical expressions for   electromagnetic
corrections to the
 lepton   anomalous magnetic momentum $a_L$
   from diagrams with inserts of   vacuum polarized operators consisting  of  closed lepton loops
formed by    leptons of the same types as $L$ and by leptons $\ell$
different from $L$.
   Having obtained explicit expressions for $a_L$, one  can
    easily  compute it
  numerically  with any predetermined accuracy and investigate    individual  contributions  from
  different diagrams of given  orders as functions of   the ratio of lepton masses of $r=m_\ell/m_L$
  with controlled accuracy.
   The presented methods are quite flexible and   versatile and can be applied in other areas where  one solves
  specific functional equations and calculates complex
integrals in analytical explicit forms.

\setcounter{equation}{0}

\section{\bf Mellin-Barnes representation}

The Mellin-Barnes representation for   massive propagators is among
the most  popular and effective approaches in analytical
investigations of multi-loop Feynman
diagrams~\cite{Kotikov:2018wxe,Mellin2,Mellin3,Boos}. It is  widely
used in calculations of  vertex  radiative corrections due to the
self-interaction of leptons with their own electromagnetic field. In
the external field $A_\mu^{(ext)}(q^2)$, the scattering amplitude
  $T(p_1,p_2)$   is determined by two form-factors, the
Dirac  $F_1(q^2)$   and Pauli  $F_2(q^2)$    normalized to
  $F_1(0)=1$, and
\begin{equation}
 F_2(0)=a_L =\left({g_L-2}\right)/2    ,\label{gminusdva}
\end{equation}
respectively. From the Dirac equation~\cite{dirac} it follows that for
a free particle with spin  $1/2$ the gyromagnetic factor $g_L$ is
exactly $2$. However, as mentioned, experimentally,  the gyromagnetic
factor  deviates form this value  $g_L\neq 2$. Thus, the gyromagnetic
factor   determines the  nonzero  $a_L\neq 0$  anomaly. Evidently,
this difference of $a_L$ from the predictions  of the free Dirac
equation is caused by the self-interaction of leptons with their own
field. Theoretically, $a_L$ is found by calculations of the
electromagnetic corrections to the lepton electromagnetic vertex and
by comparison with the definition (\ref{gminusdva}).

It should be noted  that due to the large amount of Feynman diagrams,
wholly analytical calculations of $a_L$ are quite complicated, if not
at all impossible. Nevertheless, there is a class of diagrams
consisting solely of insertions of the vacuum polarisation operators
within which calculations can be reduced to considerations of only
one Feynman diagram with the exchange of only one but massive
photon~(cf. Ref.~\cite{Lautrup:1969fr} and references therein quoted).
 For such (massive) diagrams the lepton  anomalous
 magnetic moment   is known explicitly~\cite{berestetski,brodski}
which allows one to express  the  anomaly $a_L$   as a folding
integral of the anomaly of the   "massive"  photon  with the
corresponding vacuum polarization operators, see
Refs.~\cite{Aguilar:2008qj,Solovtsova23,Solovtsova-2019,Solovtsova-Vesti,Solovtsova-2024}:
 \begin{eqnarray}
&&a_L =\frac 1\pi\int\limits_0^ \infty\frac{dt}{t}
 \Imm \widetilde\Pi(t)\frac\alpha\pi\int\limits_0^1 dx\frac{x^2 (1-x)}{x^2+(1-x)t/m_L^2}
\nonumber \\
 &&~~~~~
=
 -\frac\alpha\pi\int\limits_0^1 dx (1-x) \widetilde\Pi\left(-\frac{x^2}{1-x}m_L^2\right),~
 \label{double}
\end{eqnarray}
 where  $\widetilde\Pi $ is defined via the full virtual photon propagator   $D_{\alpha\beta}(k^2) $ as
\begin{eqnarray}
&& D_{\alpha\beta}(k^2) =  -i {g_{\alpha\beta}}\; \frac{1}{k^2} \;
\frac{1}{1+\Pi(k^2)}=
\nonumber \\
 &&
-i\frac{g_{\alpha\beta}}{k^2}\left[ 1-\left(
\Pi(k^2)-\Pi^2(k^2)+\Pi^3(k^2)+\cdots\right)\right]
\nonumber \\
 &&
\equiv -i \frac{g_{\alpha\beta}}{k^2}\left[ 1-
\widetilde\Pi(k^2)\right]. ~~~ \label{prop}
\end{eqnarray}
  In Eq.~(\ref{prop}),  the quantity $\Pi(k^2)$ represents the sum of
  all  irreducible diagrams  up to the desired order; consequently, each power of $\Pi(k^2)$
determines a  particular type of diagrams of the considered  order. In
obtaining Eq.~(\ref{double}), the dispersion relations for the photon
propagator $\Pi(k^2)/k^2$ were used.

In practice, due to  a rather complex structure of the polarization
operator (\ref{prop}),
  expression  (\ref{double}) is not  suitable enough   for numerical calculations.
Furthermore,  Eq.~(\ref{double}) does not explicitly reflect  the
physical content  of different types of diagrams and contributions of
different kinds of leptons. Consequently, a further simplification of
this expression is required for analytical  calculations.
 For this, Eq.~(\ref{double}) can be written in  a more conventional way if one applies the Mellin-Barnes representation
 to the propagator-like expression   $\frac{x^2 (1-x)}{x^2 +(1-x) t/m_L^2}$. Notice that the Mellin-Barnes representation
 is   quite popular in the literature in  multiloop calculations
 of Feynman diagrams, cf. Refs.~\cite{Kotikov:2018wxe,Mellin2,Boos}.
This representation is also  widely used in theoretical investigations
of $a_L$ as Taylor  series with respect to the ratio of the lepton
masses, see Refs.~\cite{Friot:2005cu,Aguilar:2008qj,Friot:2011ic}.

Here bellow  we briefly recall the most important aspects of the
Mellin-Barnes representation for massive photon propagators in the
context with computation of the    Feynman diagrams consisting of $n$
closed lepton loops, $n=p+j$, where  $p$ and $j$ denote the number of
loops of the leptons of    $L$ and $\ell$, respectively. For these
diagrams on has, see also Ref.~\cite{Solovtsova23},
  \begin{eqnarray}&&
 a_L{(p,j)} =
 F_{(p,j)}\frac\alpha\pi \int\limits_0^\infty\frac{dt}{t}
 \int\limits_0^1 dx \frac{x^2 (1-x)}{x^2 +(1-x) t/m_L^2}
 \nonumber \\ && \hspace*{-5mm} \!\!\!\!\!\!\!\! ~~~~~~~
\times \left[ \Pi^{(L)} \left(-\frac{x^2 }{1-x)}m_L^2 \right)\right
]^p \frac 1\pi \; \Imm \left[ \Pi^{(\ell)}(t)\right ]^j,
\label{secnd}
  \end{eqnarray}
where  $F_{(p,j)}=(-1)^{p+j+1}C_{p+j}^p$ and $C_{p+j}^p$ is a combinatoric factor.

Further, the Mellin-Barnes representation is applied to the
$x$-integration in Eq.~(\ref{secnd}). To this end, one observes that
the Euler $\beta$-function
   \begin{equation}
   B(s,\xi)=\int\limits_0^\infty dx \; \frac{x^{s-1}}{(1+x)^{s+\xi}} \,   \label{odin}
   \end{equation}
   for a specific choice of the parameters   $s$ and  $\xi$, namely for    $\xi = \beta-s$,
   \begin{equation}
   B(s,\beta-s)=\int\limits_0^\infty dx \; \frac{x^{s-1}}{(1+x)^{\beta}}=
   \int\limits_0^\infty dx \;  x^{s-1} F(x,\beta)  ,\label{dva-beta}
   \end{equation}
  is nothing but the Mellin transform of the function
   \ba
   F(x,\beta)=\displaystyle\frac{1}{(1+x)^\beta}. \label{ee}
   \ea

Then the inverse Mellin transform reads as
\begin{eqnarray}
&& \displaystyle\frac{1}{(1+x)^\beta}= \frac{1}{2\pi
i}\int\limits_{c-i\infty}^{c+i\infty}  ds \ x^{-s} \  B(s,\beta-s)
\nonumber \\
 &&~~~
  =
 \frac{1}{2\pi i}\int\limits_{c-i\infty}^{c+i\infty}  ds \  x^{-s}\
 \frac{\Gamma(s)\ \Gamma(\beta-s)}{\Gamma(\beta)} \, ,
 \label{tri-MB}
 \end{eqnarray}
where $0 <  c < \beta$. Expression  (\ref{tri-MB}) is known in the
literature as the Mellin-Barnes (MB) representation for propagator
functions. With respect to the $x$-integrand in  (\ref{secnd}),  the
MB representation leads to  the following expression for the anomalous
magnetic moment of leptons:
 \ba &&
  a_L(p,j)=\frac{\alpha}{\pi} \frac{F_{(p,j)}}{2\pi i}
\int\limits_{c-i\infty}^{c+i\infty} ds \;
 \left( \frac{4m_{\ell}^2}{m_L^2}\right)^{-s}
\nonumber \\
 &&
\times
  \Gamma(s)\Gamma(1-s)~
  \left(\frac{\alpha}{\pi}\right)^{p}\Omega_p(s)
\left(\frac{\alpha}{\pi}\right)^{j}R_j(s), \nonumber\\ &&
\label{fin1}
 \ea
where the Mellin momenta $\Omega_p(s)$ and  $R_j(s)$ are defined
through the polarization operators $\Pi^{(L)}$ and  $\Pi^{(\ell)}$
according to
\ba &&
  \hspace{-0.3cm}
\left(\frac{\alpha}{\pi}\right)^p\Omega_p(s)=
  \int_0^1 dx \; x^{2s} (1-x)^{1-s}
\nonumber \\ &&
 ~~~\times \left[ \Pi^{(L)} \left( -\frac{x^2 }{1-x}m_L^2  \right)
  \right ]^p , \label{Omp}
  \\[0.2cm] &&
  \hspace{-0.3cm}
\left(\frac{\alpha}{\pi}\right)^j R_j(s)= \int_0^\infty
\frac{dt}{t}\left( \frac{4m_{\ell}^2}{t}\right)^s \frac 1\pi \Imm
\left[ \Pi^{(\ell)} (t)\right ]^j \label{Rj}. \ea
In Eqs.~(\ref{Omp}) and  (\ref{Rj}) the polarization operator
$\Pi^{(L,\ell)}$
 is well-known explicitly: for the second order "``bubble''\,-like  diagrams
their explicit form can be found in, e.g. Ref.~\cite{Lautrup:1969fr}:
\begin{eqnarray}&& \Ree \; \Pi^{(L,\ell)} (q^2) = \left(\frac{\alpha}{\pi}\right
) \left[\frac89 - \frac{\delta^2}{3} \right.
\nonumber \\
 &&~~~~~
\left. ~~~~~~~~~~~~ +\delta \left(\frac12
-\frac{\delta^2}{6}\right)\ln\frac{|1-\delta|}{1+\delta}\right],
\label{reP1} \\[0.2cm]
&&
\frac1\pi \Imm  \;  \Pi^{(L,\ell)}(q^2) =
\left(\frac{\alpha}{\pi}\right ) \delta \left(\frac12 -\frac16
\delta^2\right)
\nonumber \\
 &&~~~~~~~~~~~~~~~~~\times
\theta\left (q^2-4m^2_{(L,\ell)}\right), \label{imP1}
\end{eqnarray}
where $\delta=\sqrt{1-4m^2_{(L,\ell)}/{q^2}}$.
 Notice that since the dependence of the operators (\ref{reP1}) and (\ref{imP1}) on  $q^2$
 enters via the   combination  $m_L^2/q^2$,  the operator
 $\Pi^{(L)} \left( -\frac{x^2 }{1-x}m_L^2 \right)$  depends only on  $x$,
\ba && \Pi^{(L)} \left( -\frac{x^2 }{1-x}m_L^2
\right)=\frac\alpha\pi \left [\frac59 +\frac{4}{3x} - \frac{4}{3x^2}
\right.
\nonumber \\
 &&~~~~~
\left. + \left (-\frac13 +\frac{2}{x^2}-\frac{4}{3x^3}\right )
\ln(1-x)\right]. \label{lashk}
 \ea

   For irreducible diagrams of the mixed type (lepton loop crossed by a photon line)
 the polarizatioln operators (of the fourth order) are of a more complicated
   structure and can be found in Res.~\cite{Lautrup:1969fr,sabry,Laursen:1978nw}.
Equations (\ref{fin1})-(\ref{imP1}) demonstrate that the anomalous
magnetic moment
 $a_L$ is entirely determined by powers of the polarization operator  $\Pi(t)$ that can lead to rather
 cumbersome  expressions for   $\Omega_p(s)$ and $R_j(s)$. This requires specific methods and techniques for
 their analytical calculations.

\section{\bf The moment integrals }

We are  now left with two types of moments: $\Omega_p$ and $R_j$. The
first one, due the Euclidean nature of its argument, depends only on
the real part of the polarization  operator and can be computed
straightforwardly by decomposing the logarithm $\ln(1-x)$ into the
Taylor series and integrating the resulting expression term by term
(cf. also Ref.~\cite{Aguilar:2008qj}). More complicated integrals
originate from the moments $R_j(s)$ which involve both the real and
imaginary parts  of the polarization operator. For $j\ge 3$, the
typical integrals requiring special attention read as, cf.
Ref.~\cite{Aguilar:2008qj}),
 \ba
  \Phi_n(s)=\int\limits_0^1 dx (1-x^2)^s\ln^n\left(\frac{1-x}{1+x}\right), \label{integral}
 \ea
 which  is a corollary  of insertions of the
    polarization operators (\ref{reP1}) and  (\ref{imP1}) in to  $R_j$, Eq.~(\ref{Rj}).
 Regardless this integral can be expressed explicitly in terms of   hypergeometric  functions as
 \ba&&
\Phi_n(s)=\frac{(-1)^n 2^{2s+1}n!}{(s+1)^{n+1}} \times
 \\
 && {}_{n+2}F_{n+1}
\left(\begin{matrix}s+1&s+1 &s+1&\cdots& 2s+ 2
\\ s+2&s+2&s+2&\cdots&\end{matrix}\middle| -1  \right), ~~~~
\nonumber
\label{phin} \ea
  the   further usage  of Eq.~(\ref{phin}) in  computation of contour integrals
   by the Cauchy theorem is highly hindered. This is because  the integration variable
$s$ in Eq.~(\ref{fin1}) is rather  a parameter  of   the
hypergeometric function   than a true integration variable.
 Therefore, it is tempting to express $  {}_{n+2}F_{n+1} (\ldots )$
 in terms of superpositions
 of functions of only one variable~$s$. There are few possibilities  to achieve this.

  \subsection{\bf Functional equation.   }

  Integration (\ref{integral}) by parts leads to the following recursive
  functional equation for $\Phi_n(s) $:
\ba
(1+2s) \Phi_n(s) &-& 2s \Phi_n(s-1)
\nonumber \\[0.2cm]
 &+&
\frac{2n(n-1)}{s}
\Phi_{n-2}(s-1)=0,
 \label{funceq1} \ea which, in principle, allows one
to express the integrals
 $\Phi_n(s)$ for any value of $n \geq 2 $ and any $s \geq 1$
 in terms of special functions. However, it should be noted that there is no universal algorithm  to solve
 in general  functional equations.
Each  type of equations requires individual approach in seeking for a
solution. In our case, one observes that for   $n=2$
Eq.~(\ref{funceq1}) reduces to
 (cf. also Ref.~\cite{Aguilar:2008qj})

 \begin{equation}
 (1+2s) \Phi_2(s) - 2s\Phi_2(s -1) +\frac{2\sqrt{\pi}}{s} \frac{\Gamma(s)}{\Gamma(1/2+s)}=0
\label{funceqn22}
 \end{equation}
with the initial conditions
$\Phi_2(0)=\pi^2/3$. The free term in Eq.~(\ref{funceqn22}) suggests
that the solution of Eq.~(\ref{funceqn22}) should be looked for as
\ba \Phi_2(s)=\frac{\sqrt{\pi}\Gamma(1+s)}{\Gamma(3/2+s)} {\cal
F}_2(s+1). \label{calF} \ea Inserting  Eq.~(\ref{calF}) in
Eq.~(\ref{funceqn22}) we find that the auxiliary function
 ${\cal F}_2(s) $ obeys the recurrent relation of the type
\ba {\cal F}_2(s+1)-{\cal F}_2(s) +\frac{1}{s^2}=0, \label{rec} \ea
which exactly coincides with the corresponding recurrent relation for
the polygamma  function  $\Psi^{(1)}(s)$ of the first order.
Thus, for
$n=2$ the solution of the functional equation (\ref{funceqn22}) reads
as
\ba \Phi_2(s) = \frac{\Gamma(s+1)}{\Gamma(s+3/2)}
\sqrt{\pi}\Psi^{(1)}(s+1). \label{phi2} \ea
  Now we are in a position  to find  $\Phi_n(s)$ for any even  value of  $n$  by starting
with  $n=2$ and consecutively increasing  $n$ keeping it even and
using Eqs.~(\ref{funceq1}) and ~(\ref{funceqn22}). So for  $n=4$ one
has \ba
 \Phi_4(s)& = &\frac{\sqrt{\pi}\Gamma(s+1)}{\Gamma(3/2+s)}
 \nonumber\\ & 
 \times &\left[\left(6\Psi^{(1)}(1+s)\right)^2+\Psi^{(3)}(1+s)\right]. \label{phi4} \ea
For  $n=6$ the solution is \ba
\Phi_6(s)&=&\frac{\sqrt{\pi}\Gamma(s+1)}{\Gamma(3/2+s)}\left[\Psi^{(5)}(1+s)+60
\left(\Psi^{(1)}(1+s)\right)^3 \right.
\nonumber \\
 &+& \left.  30\Psi^{(3)}(1+s)\Psi^{(1)}(1+s)\right] .
\label{phi61} \ea For odd $n$ one should use the explicit form of the
hypergeometric function in terms of the Pochammer symbols.

\subsection{\bf The  PSLQ-algorithm.\, }

Another  more sophisticated way of expressing the hypergeometric
function (\ref{phin}) through the polygamma functions consists in
exploiting the so-called PSLQ algorithm of finding an integer
relation between two vectors, recently discovered by H.~Ferguson and
D.~Bailey~\cite{Ferguson1,Ferguson}. The gist of the method is   the
following: let $X(x_1, x_2, \cdots, x_n)$ be a vector of real
components $x_i$.  Then $X$  is said to possess an integer relation
if there exists a vector $A(a_1, a_2, \cdots  a_n) $ with integers
$a_i$ not all zero, such that $(A\cdot X)=0$, i.e.
 \ba \left( A \cdot  X\right )= a_1 x_1 + a_2 x_2 \cdots  a_n x_n =0.
\label{orto} 
\ea
 In this case, relation (\ref{orto})  allows one to
express one of the components of $X$, say $x_1$,
 as a linear combination of the remaining $x_2\ldots x_n$ components.

 The integer relation proves particularly   effective in the case when some preliminary information on the   nature of $x_i$
 is available, for instance,  if  $x_2\ldots x_n$ are the results of computation of some special functions or powers
 of $\pi$ and logarithms $\ln$, etc. Then evidently,
$x_1$ can be expressed as a superposition of these  powers of $\pi$,
logarithms and the form of some additional relations with special
functions. In our case, we define  $x_1$ as a solution of
Eq.~(\ref{phin}) at some (arbitrary) value of $s$. Additional
information on  $x_2\ldots x_n$ comes from the fact that at   $n$ even
the solution of Eq.~(\ref{phin}) is a linear combination of
 $\pi^n$, $\pi^{n-2}$, $\pi^{n-4},$...$\pi^0,$ with coefficients depending on  $s$.
On the other side, Eq.~(\ref{phi2}) can serve as a hint  that the
solution of Eq.~ (\ref{phin}) also
 contains polygamma functions of the odd order
 with the argument  $s+1$.
 Then, for instance, for  $n=6$, the vector  $X$ consists of all   possible
 combinations of polygamma functions  which provide powers of $\pi$
up to $\pi^6$. Taking into account Eq.~(\ref{calF}), one defines
\ba
 &&x_1=\Phi_6(s), \nonumber \\
 &&  x_2=\frac{\sqrt{\pi} \Gamma(s+1)|}{\Gamma(3/2+s)}\big(\Psi^{(1)}( s + 1)\big)^3,\label{vector} \\ &&
x_3= \frac{\sqrt{\pi} \Gamma(s+1)|}{\Gamma(3/2+s)}   \Psi^{(1)}(s +
1)\Psi^{(3)}( s + 1), \,
\nonumber \\
 &&
 x_4=\frac{\sqrt{\pi}
\Gamma(s+1)|}{\Gamma(3/2+s)}\Psi^{(5)}( s + 1).\nonumber
\ea

With this definition, the vector  $X=[x_1,x_2,\ldots x_n]$ serves as
input for the  PSLQ algorithm, which is implemented as standard
subroutines   in libraries of symbol manipulation packages,  e.g.,
such as  Wolfram Mathematica or Waterloo Maple. The run of such
packages for the PSLQ algorithm  (for  $n=6$) provides the vector
$A=(1, -60, -30, -1)$, which reproduces exactly the
solution~(\ref{phi61}).

Analogously, for $n=8$ the vector $X$ is defined as \ba
 &&x_1=\Phi_8(s), \quad  x_2=\frac{\sqrt{\pi} \Gamma(s+1)}{\Gamma(3/2+s)} \Psi^{(7)}( s + 1); \nonumber \\ &&
 x_3=\frac{\sqrt{\pi} \Gamma(s+1)}{\Gamma(3/2+s)} \Psi^{(5)}( s + 1) \Psi^{(1)}( s + 1);
  \nonumber \\
  &&
x_4=\frac{\sqrt{\pi} \Gamma(s+1)}{\Gamma(3/2+s)} \big( \Psi^{(3)}( s +
1) ( s + 1)\big)^2  ;  \nonumber \\ && x_5=\frac{\sqrt{\pi}
\Gamma(s+1)}{\Gamma(3/2+s)} \big( \Psi^{(1)}( s + 1)\big)^2\Psi^{(3)}(
s + 1);\nonumber \\ && x_6=\big( \Psi^{(1)}( s + 1)\big)^4.
\label{vector1} \ea The PSLQ run provides $A=[-1,1,56,70,840,840]$,
i.e.,
 \ba && \Phi_8(s) = \frac{\sqrt{\pi} \Gamma(s+1)}{\Gamma(3/2+s)}
\left\{
 \Psi^{(7)}( s + 1) + 56\Psi\Psi^{(5)}\right.\nonumber \\ && \left .
 \times( s + 1)\Psi^{(1)}( s + 1)
 +70 \big(\Psi^{(3)}( s + 1)\big)^2+
 840
\right.\nonumber \\ && \left .
 \times\big( \Psi^{(1)}( s + 1)\big)^2 \Psi^{(3)}( s + 1)
\right.\nonumber \\ && \left .
 + 840\big( \Psi^{(1)}( s + 1)\big)^4 \right\}.
 \ea

\subsection{\bf Analytical continuations of the hypergeometric functions.}

The computation of the Mellin moments gives rise to hypergeometric
functions of the type
$_3F_2\left(\frac{1}{2},\frac{1}{2},1;\frac{3}{2},\frac{3}{2}+s;1\right)$,
for which the integral  representation reads as
\begin{equation}
 {_3}F{_2}\left(\frac12,\frac12,1;\frac32,s+\frac32;1\right)=
  -\frac{1}{\sqrt{\pi}}  \displaystyle\frac{\Gamma\left(s+\frac32\right)}{\Gamma\left(s+1\right)}\;{ \cal  I}(s),
\label{resder}
\end{equation}
where
\begin{equation}
{ \cal I}(s)\equiv\int\limits_0^1\frac{dy}{y}
(1-y^2)^{s}\ln\bigg[\displaystyle\frac{1-y}{1+y}\bigg]. \label{Iz}
\end{equation}
It can be seen that due to the presence of   $\Gamma$-functions in
Eq.~(\ref{resder}), the hypergeometric function is defined only in the
region $ \Ree  s>-1$. This makes problematic the use of the Cauchy
theorem  in the left semi-plan  $ \Ree s  <-1 $. Moreover, the
integral (\ref{Iz}) does not converge for negative integer  $s$. Thus,
for further calculations  an analytical continuation of
$_3F_2\left(\frac{1}{2},\frac{1}{2},1;\frac{3}{2},\frac{3}{2}+s;1\right)$
 in the whole complex plane of $s$  is necessary.
For this  purpose, we derive a functional equation for the integral  $
{\cal I}(s)$ by writing the integrand in  (\ref{Iz})  as
\begin{eqnarray}&&
(1-y^2)^s\frac{1}{y}\ln\left(\frac{1-y}{1+y}\right)=(1-y^2)^{s-1}\frac{1}{y}\ln\left(\frac{1-y}{1+y}\right)
\nonumber \\
 &&~~~~~
-y(1-y^2)^{s-1}\ln\left(\frac{1-y}{1+y}\right). \label{Iz-2}
\end{eqnarray}
The first term  in (\ref{Iz-2})  defines the integral ${\cal I}(s-1)$;
the second one, after integration by parts, reads as
\begin{eqnarray}&&
 -\int\limits_{0}^{1}y(1-y^2)^{s-1}\ln\left(\frac{1-y}{1+y}\right)\;dy
\nonumber \\
 &&
 =
\int\limits_{0}^{1}\frac{1}{s}(1-y^2)^{s-1}\;dy=\frac{\sqrt{\pi}}{2s}\frac{\Gamma(s)}
{\Gamma\left(\frac{1}{2}+s\right)}\;. \label{e231}
\end{eqnarray}

In this way one obtains the functional equation for
  $ {\cal I}(s)$ in the following form:
\ba \label{Eq-Iz} {\cal I}(s)={\cal I}(s-1)+
\frac{\sqrt{\pi}}{2s}\frac{\Gamma(s)}
{\Gamma\left(\frac{1}{2}+s\right)}\;. \label{gz} \ea
 The solution of this equation for integer $s=n$ is
\begin{equation}
{\cal I}(n)={\cal I}(0)+\sum\limits_{k=1}^{n}
\frac{\sqrt{\pi}}{2k}\frac{\Gamma(k)}
{\Gamma\left(\frac{1}{2}+k\right)}\;,
\end{equation}
where ${\cal I}(0)=-\frac{\pi^2}{4}$, and the above  sum   is related
to the hypergeometric function as
\begin{eqnarray} &&
\sum\limits_{k=1}^{n} \frac{\sqrt{\pi}}{2k}\frac{\Gamma(k)}
{\Gamma\left(\frac{1}{2}+k\right)}=\frac{\sqrt{\pi}\Gamma(n+1)}{2(n+1)\Gamma\left(\frac{3}{2}+n\right)}
\nonumber \\
 &&\times
{_3}F{_2}
\left(1,1+n,1+n;\frac{3}{2}+n,2+n;1\right)+\frac{\pi^2}{4}\; .
\end{eqnarray}

This implies that
\begin{equation}
\label{hyper} {\cal I}(n)=-\frac{\sqrt{\pi}\Gamma(n+1){_3}F{_2}
\left(1,1+n,1+n;\frac{3}{2}+n,2+n;1\right)}{2(1+n)\Gamma\left(\frac{3}{2}+n\right)}\;.
\end{equation}

Coming back    to the variable   $s$,  the desired analytical
continuation reads as
\begin{equation}
\label{Iz-new} {\cal I}(s)=-\frac{\sqrt{\pi}\Gamma(s+1){_3}F{_2}
\left(1,1+s,1+s;\frac{3}{2}+s,2+s;1\right)}{2(1+s)\Gamma\left(\frac{3}{2}+s\right)}\;.
\end{equation}

Hence, for  $\Ree(s)>-1$ the following equality holds:
\begin{eqnarray} &&
{_3}F{_2}\left(\frac{1}{2},\frac{1}{2},1;\frac{3}{2},\frac{3}{2}+s;1\right)
\nonumber \\[0.2cm]
 &&
=\frac{{_3}F{_2}\left(1,1+s,1+s;\frac{3}{2}+s,2+s;1\right)}{2(s+1)}\;.
\label{3F2-new-3}
\end{eqnarray}

Since the hypergeometric function in the r.h.s. is well defined in the
whole complex plane of $s$ (with poles at $s=-n$),
 the sought analytical continuation is
\begin{eqnarray}&&
{_3}F{_2}\left(\frac{1}{2},\frac{1}{2},1;\frac{3}{2},\frac{3}{2}+s;1\right)
\Bigg|_{\Ree (s)<-1} \Longrightarrow \nonumber  \\[0.2cm]
 &&
\frac{{_3}F{_2}\left(1,1+s,1+s;\frac{3}{2}+s,2+s;1\right)}{2(s+1)}\;.
\label{3F2-new-3bis}
\end{eqnarray}
Analogous manipulations with  another hypergeometric function
frequently arising in our calculations, $
{_3}F{_2}\Bigl(\frac{1}{2},\frac{1}{2},1;\frac{3}{2},$
$\frac{7}{2}+s;1 \Bigr) $  (which is valid for $\Ree (s) >-3$) lead to
the   replacement
\begin{eqnarray}&&
{_3}F{_2}\left(\frac{1}{2},\frac{1}{2},1;\frac{3}{2},\frac{7}{2}+s;1\right)\Bigg|_{\Ree (s) <-3}
\Rightarrow \nonumber  \\[0.2cm]
 &&
\frac{{_3}F{_2}\left(1,3+s,3+s;s+4,\frac{7}{2}+s;1\right)}{2(s+3)}\;.\label{3F2-new-7}
\end{eqnarray}

Thereby, equations (\ref{3F2-new-3}) and  (\ref{3F2-new-7}) represent
the analytical continuation of the relevant hypergeometric functions
in the whole complex plane of $s$.

\subsection{\bf Integrals of
the type $\int_0^1 (1-y^2)^s\frac1y \ln^2(1-y^2)
\ln\left(\frac{1-y}{1+y}\right) dy$.}\,

\vspace*{0.1cm}

Another integral, frequently appearing in our concrete calculations is
the  following
 \ba I(s) = \int_0^1 (1-y^2)^s\frac1y \ln^2(1-y^2)
\ln\left(\frac{1-y}{1+y}\right) dy, \label{In} \ea
 which is nothing
but the second derivative of (\ref{Iz}).
 Consequently, the second derivative of  Eq.~(\ref{gz}) leads  to the functional equation
 \ba
 I(s+1) -I(s) =\frac{d^2 g(s)}{ds^2} , \label{fun}
 \ea
 where (cf. also Eqs.~ (\ref{e231})), (\ref{gz}) )
\ba  && \hspace*{-0.3cm} g(s)= -\int_0^1 y(1-y^2)^s
\ln\left(\frac{1-y}{1+y}\right)dy
\nonumber \\
 &&
 \hspace*{-0.3cm}
=\frac{1}{2(s+1)} B\left(\frac12,s+1\right)=
\frac{\sqrt{\pi}}{2(s+1)}\frac{\Gamma(s+1)}
{\Gamma\left(\frac{3}{2}+s\right)}.  \label{gs} \ea where
$B\left(\frac12,s+1\right)$ denotes the  Euler $\beta$-function.
 The last two relations  resulted from integration by parts  in Eq.~(\ref{gs})  by performing a
 change of variables
  $u= \ln\left(\frac{1-y}{1+y}\right)$ and
   $ v=(1-y^2)^{s+1}/2/(s+1).$

For integer  $s=n\ge 1$ the solution of Eq.~(\ref{fun}) is \ba
I(n)=I(0) + \sum\limits_{k=0}^{n-1 }    g''(s)\bigg|_{s=k}.
\label{Ieq} \ea where \ba &&
 I(0) =  \int_0^1 \frac1y \ln^2(1-y^2)
  \ln\left(\frac{1-y}{1+y}\right) dy
\nonumber \\
 &&
   =2\int\limits_0^1\frac{\ln(\xi)}{1-\xi^2} \ln^2\left(\frac{4\xi}{(1+\xi)^2}\right)   d\xi. \label{pervyi}
  \ea
  Attempts to resolve  $I(0)$ in quadratures show that the resulting expression
 contains
$\zeta(3)$, ${\rm Li}_4(1/2) $ and powers of  $\pi$ and $\ln(2)$.
However, the exact expression for the integral cannot be obtained by
direct integration. To avoid cumbersome calculations, we again employ
the PSLQ-algorithm.
 This time, the input vector is chosen as
\ba && X  =   \left[I_{calc.},
  \pi^4,  \ln(2),  \ln^2(2),  \pi^2 \ln^2(2),
\right.
\nonumber \\
 &&~~~~~
\left.
    \ln^4(2), {\rm Li}_4(1/2),  \ln(2)\zeta(3)\right],
\ea
where $I_{calc.}$ is the result of high-precision numerical
calculation of $(\ref{pervyi})$. Then the PSLQ routine provides
$(A\cdot X)=0$ with
$$A=[-360, -49, 0, 0, -120, 120, 2880, 2520].$$

Finally, \ba
 && I(0)= 2\int\limits_0^1\frac{\ln(\xi)}{1-\xi^2} \ln^2\left(\frac{4\xi}{(1+\xi)^2}\right)   d\xi=
 - \frac{49}{360}\pi^4   - \frac13 \nonumber
  \\   &&\times\pi^2\ln^2(2) +\frac13  \ln^4(2)+8 {\rm Li}_4(1/2)
  + 7  \zeta(3)\ln(2)).
  \label{io}
\ea

\subsubsection
  {\bf Calculation of the derivative  $\mathbf{  g''(s) }$  in (\ref{gs}) - (\ref{Ieq}).}

 With the result~(\ref{io}) we are now left with only calculations   of  the derivative
$  g''(s)  $  in Eqs.~(\ref{gs}) - (\ref{Ieq}). For this purpose, one
can use a useful method of calculations of derivatives of complex
expressions by
 introducing  the logarithmic derivative  $ L_g (s)=g'(s)/g(s)$. Then
\ba \frac{\partial^2 g(s)}{\partial s^2} = g(s) \left[ L_g^2(s) +
L'_g(s)\right]. \label{diff} \ea In our case $g(s)$ is defined by
Eqs.~(\ref{gs}), so that
 \ba
  && L_g(s) = \Psi(s+1)-\Psi(s+3/2)-\frac{1}{s+1},\\
   &&L_g' (s)= \Psi(1,s+1)-\Psi(1,s+3/2)+\frac{1}{(s+1)^2}. \nonumber
\ea 
Eventually, the integral  (\ref{Ieq}) can written in the form \ba
I(n)=I(0) +\sum_{k=0}^{n-1} S_k ,\label{res} \ea where
 \vspace*{-0.3cm}
 \ba &&
 \hspace*{-0.3cm} S_k=
  \frac{1}{2(k+1)} B\left(\frac12,k+1\right)
 \nonumber \\
 &&
  \times \Bigg[
  \left(\Psi(k+1)-\Psi(k+3/2)-\frac{1}{k+1}\right)^2  \nonumber \\ &&
 +\Bigg(\Psi(1,k+1)
 -\Psi(1,k+3/2)+\frac{1}{(k+1)^2}\Bigg)\Bigg] .
\ea
 Now it is straightforward to compute ~(\ref{In})  for any value of $n$.

For instance, $I(0) = -4.789892682$, $S_0=4.314177755$,
$S_1=0.337852415$, $S_2=0.0781477163$, $S_3=0.0278969038$  \ldots The
integrals themselves  are:

 \vskip 1mm\noindent
  $I(1)= I(0)+S_0 = -0.475714927,$\\[1mm]
 $I(2) = I(0)+S_0 +S_1= -0.137862513,$\\[1mm]
 $I(3) = I(0)+S_0 +S_1+S_2=-0.059714796,$\\[1mm]
 $I(4) =  I(0)+S_0 +S_1+S_2+S_3=-0.031817892, \, etc.$

\section{\bf  Summary}

We have presented some of the most typical methods and techniques used
in
our~\cite{Solovtsova23,Solovtsova-2019,Solovtsova-Vesti,Solovtsova-2024}
analytical calculations of   radiative corrections of high order (up
to the sixth order in the fine structure constant $\alpha$)
 to the anomalous magnetic moment of leptons.
 It is demonstrated that quite complex integrals containing powers of logarithm
 can be reduced to functional equations which are solved by using
 the relatively new PSLQ-algorithm of finding  integer relations between two vectors.
 The functional equation method is also used in the procedure of analytical continuation
 of some specific hypergeometric  functions entering into our computations.

The presented methods are rather general and can be applied
successfully in many fields when computing analytically complex
integrals which result  in different special functions, e.g.,
polygamma functions, polylogarithms, Riemann's zeta functions etc.
 \section{Acknowledgments}
 We are very grateful to V. Bytiev and A. Kotikov for fruitful discussions on the properties of the hypergeometric functions,
 PSLQ-algorithm and Mellin-Barnes representation.
 The performed investigations were supported by a grant of cooperation between Republic of the  Belarus and the
 Joint Institute for Nuclear Reactions.


\begin{thebibliography}{100}

\bibitem{dirac}              
P.~A.~M.~Dirac, The quantum theory of the electron,
               Proc. Roy. Soc. Lond. A {\bf 117}, 610 (1928).
\bibitem{Parker:2018vye}
R.~H.~Parker, Y.~Chenghui, W.~Zhong, B.~Estey, H.~Muller, Measurement
of the fine-structure constant as a test of the Standard Model. {
Science} {\bf 360}, 191 (2018).

\bibitem{Morel}
L.~Morel, Z.~Yao, P.~Clad,  Guellati-Khelifa S. Determination of the
fine-structure constant with an accuracy of 81 parts per trillion, {
Nature} {\bf 588}, 61 (2020).

\bibitem{Fan:2022eto}
                       X.~Fan, T.~G.~Myers, B.~A.~D.~Sukra and G.~Gabrielse,
                       Measurement of the electron magnetic moment,
                       Phys. Rev. Lett. \textbf{130}, 071801 (2023).

\bibitem{Fermilab2023}
 D.~P.~Aguillard {\it et al.},  (Muon $g - 2$ Collaboration), Measurement of
the positive muon anomalous magnetic moment to 0.127 ppb,
 {Phys. Rev. Lett.} {\bf 135}, 101802 (2025).
\bibitem{Jegerlehner:2017gek}
F.~Jegerlehner, The anomalous magnetic moment of the muon, { Springer
Tracts in Modern Physics}, 1 (2017).

\bibitem{review-2025}
R.~Aliberti {\it et al.}, The anomalous magnetic moment of the muon in the
Standard Model: an update, { Phys.\ Rept.} {\bf 1143}, 1 (2025).

\bibitem{CMD-3:2023alj}
F.~V.~Ignatov {\it et al.}, Measurement of
 the $e^+e^-\to\pi^+\pi^-$ cross section from threshold to 1.2 GeV with the
 CMD-3 detector, Phys. Rev. D {\bf 109}, 112002 (2024).

\bibitem{CMD-3:2023rfe}
F.~V.~Ignatov {\it et al.}, Measurement of the pion formfactor with
CMD-3 detector and its implication to the hadronic
 contribution to muon $(g-2)$,
    Phys. Rev. Lett. {\bf 132}, 231903 (2024).

\bibitem{Davier-2023-1}
M.~Davier, Z.~Fodor, A.~Gerardin, L.~Lellouch, B.~Malaescu,
F.~M.~Stokes
 {\it et al.}, Hadronic vacuum polarization: comparing lattice
QCD and data-driven results in systematically improvable ways,
Phys. Rev. D {\bf 109}, 076019 (2024).

\bibitem{Davier-2023-2}
M.~Davier, A.~Hoecker, A.~M.~Lutz, B.~Malaescu, Z.~Zhang,
 Tensions in $e^+e^- \rightarrow \pi^+\pi^-(\gamma)$
measurements: the new landscape of data-driven hadronic vacuum
polarization predictions for the muon $g_\mu-2$,
Eur. Phys. J. C {\bf 84}, 721 (2024)
\bibitem{Masjuan-2024}
P.~Masjuan, P.~Roig,  Data-driven approximations to the Hadronic
Light-by-Light scattering contribution to the muon (g-2),
Nuov. Cim. C {\bf 47}, 246 (2024).

\bibitem{Bryzgalov-2024}
V. Bryzgalov, O. Zenin,  A comment on the impact of CMD-3 $e^+e^-
\rightarrow \pi^+\pi^-$ cross section measurement on the SM
$g_\mu-2$ value, 
{ E-Print: arXiv:2401.07204 [hep-ph]}, 2024.

\bibitem{Volkov-2024}
S.~Volkov, Calculation of the total 10-th order QED contribution to
the lepton magnetic moments, Phys. Rev. D  {\bf 110}, 036001 (2024).

\bibitem{Laporta:2017okg}      
S. Laporta, High-precision calculations of the 4-loop contribution to the electron $g-2$ in QED,
                 Phys. Lett. B {\bf 772}, 332 (2017).

\bibitem{elliptic1}           
S.~Laporta, High-precision calculation of the 4-loop QED contribution
to the slope of the Dirac form factor, Phys. Lett. B   {\bf
800}, 135 (2020).

 \bibitem{Volkov-22}
S.~Volkov, A Method of fast calculaion of lepton magnetic moments in
quantum electrodynamics, Phys. Part. Nucl. {\bf 53}, 805 (2022).


\bibitem{Volkov-2023}
 S.~Volkov, Calculation of
lepton magnetic moments in quantum electrodynamics: A justification of
the flexible divergence elimination method, Phys. Rev. D. {\bf 109},
03007 (2024).

\bibitem{Aguilar:2008qj}      
J.~P. Aguilar, E.~de~Rafael, D.~Greynat, Muon anomaly from lepton
vacuum polarization and the Mellin-Barnes representation, Phys. Rev. D
{\bf 77}, 093010 (2008).


\bibitem{Solovtsova23}       
O.~P.~Solovtsova, V.~I.~Lashkevich, L.~P.~Kaptari,  Lepton anomaly
from QED diagrams with vacuum polarization insertions within the
Mellin-Barnes representation,  Eur. Phys. J. Plus   {\bf  138}, 212
(2023).

\bibitem{Mellin-book}
I.~Dubovyk, J.~Gluza , G.~Somogyi, Mellin-Barnes Integrals: A Primer
on Particle Physics Applications, { Lect. Notes  Phys.}  {\bf 1008},
208 (2022).

\bibitem{Boos}
E.~E. Boos, A.~E. Davydychev, A method of evaluation massive Feynman
diagrams, Theor. Math. Phys. {\bf 89}, 1052 (1991).


\bibitem{Friot:2005cu}
J.P. Aguilar, D. Greynat, E. de Rafael, Asymptotics of Feynman
diagrams and the Mellin-Barnes representation, Phys.\ Lett.\ B {\bf
628}, 73 (2005).

\bibitem{Solovtsova-2019}
 O.~P. Solovtsova, V.~I. Lashkevich, A.~V.  Sidorov,      
Some analytic results for the contribution to the anomalous magnetic
moments of leptons due to the polarization of vacuum via lepton loops,
Eur.~Phys.~J. Web Conf. {\bf 222}, 03007 (2019).


\bibitem{Solovtsova-Vesti}
V.~I. Lashkevich, O.~P. Solovtsova,  L.~P. Kaptari, Analytical
calculations of fifth-order electromagnetic corrections to the
anomalous magnetic moment of lepton within the Mellin-Barnes
representation. Proc. of the National Academy of Sciences of
Belarus   {\bf  59}, 338 (2023).


\bibitem{Solovtsova-2024}
O.~P. Solovtsova, V.~I. Lashkevich, L.~P. Kaptari,  Analytical
calculations of the  tenth order QED radiative corrections to lepton
anomalies within the Mellin--Barnes representation,
J.~Phys. G \ {\bf 51}, 055001 (2024).


\bibitem{Kotikov:2018wxe}
A.~V. Kotikov, S.~Teber,
  Multi-loop techniques for massless Feynman diagram
 calculations, Phys. Part. Nucl.  {\bf  50}, 1 (2019).


\bibitem{Mellin2}
V.~A. Smirnov,
 Analytic tools for Feynman integrals,
Springer Tracts Mod. Phys. {\bf  250}, 1 (2012).

\bibitem{Mellin3}
E.~E. Boos, A.~E.   Davydychev,            
 A method of evaluation massive Feynman diagrams,
Phys. Rev. {\bf  89}, 1052 (1991).


\bibitem{Lautrup:1969fr}
B.~E. Lautrup, E. de Rafael, Calculation of the sixth-order
contribution from the fourth-order vacuum polarization to the
difference of the anomalous magnetic moments of muon and electron,
Phys.\ Rev.\ {\bf  174}, 1835 (1968).


\bibitem{berestetski}
V.~B. Berestetskii,
O.~N. Krohnin, A.~K.  Khlebnikov,      
Concerning the radiative corrections to the $\mu$-meson magnetic
moment, Zh. Eksp. Teor. Fiz. {\bf  30}, 788 (1956).


\bibitem{brodski}
S.~J. Brodsky, E.~Rafael,      
Suggested boson-lepton pair coupling and the anamalous magnetic moment
of the muon, Phys. Rev. {\bf  168}, 1620 (1968).


\bibitem{Friot:2011ic}
S. Friot, D. Greynat, On convergent series representations of
Mellin-Barnes integrals, J.\ Math.\ Phys.\ {\bf  53}, 023508 (2012).


\bibitem{sabry}
G. K\"allen , A. Sabry,  Fourth order vacuum polarization, Dan. Mat.
Fys. Medd. {\bf  29}, 556 (1955).


\bibitem{Laursen:1978nw}
M. Laursen, M.~A. Samuel,  Corrections to the sixth order anomalous
magnetic moment of the muon, Phys. Rev. D {\bf  19}, 1281. (1979).

\bibitem{Ferguson1}
 H. R. P. Ferguson, D. H. Bailey,  S. Arno,
 Analysis of PSLQ, an integer relation finding algorithm,
 Math. Comp., {\bf 68}, 351, (1999).

\bibitem{Ferguson}
D. H. Bailey, D. J. Broadhurst,  Parallel integer relation
           detection: techniques and applications,
Math. Comp. {\bf 70}, 1719, (2000).

\end{thebibliography}
\end{document}